\documentclass{ijclclp}
\title{The Representational Alignment Hypothesis:\\
Evidence for and Consequences of Invariant Semantic Structure Across Embedding Modalities}
\author{Akhil Ramidi and Kevin Scharp}

\makeatletter
\newcommand\blfootnote[1]{%
  \begingroup
  \renewcommand\thefootnote{}%
  \renewcommand\@makefntext[1]{\noindent##1}%
  \footnotetext{#1}%
  \endgroup
}
\makeatother

\begin{document}
\maketitle
\blfootnote{Acknowledgements redacted.}
\thispagestyle{firstpage}

\begin{abstract}
There is growing evidence that independently trained AI systems come to represent the world in the same way. In other words, independently trained embeddings from text, vision, audio, and neural signals share an underlying geometry. We call this the \textit{Representational Alignment Hypothesis} (RAH) and investigate evidence for and consequences of this claim. The evidence is of two kinds: (i) internal structure comparison techniques, such as representational similarity analysis and topological data analysis, reveal matching relational patterns across modalities without explicit mapping; and (ii) methods based on cross-modal embedding alignment, which learn mappings between representation spaces, show that simple linear transformations can bring different embedding spaces into close correspondence, suggesting near-isomorphism. Taken together, the evidence suggests that, even after controlling for trivial commonalities inherent in standard data preprocessing and embedding procedures, a robust structural correspondence persists, hinting at an underlying organizational principle. Some have argued that this result shows that the shared structure is getting at a fundamental, Platonic level of reality. We argue that this conclusion is unjustified. Moreover, we aim to give the idea an alternative philosophical home, rooted in contemporary \textit{metasemantics} (i.e., theories of what makes a representation and what makes something meaningful) and responses to the \textit{symbol grounding problem}. We conclude by considering the scope of the RAH and proposing new ways of distinguishing semantic structures that are genuinely invariant from those that inevitably arise due to the fact that all our data is generated under human-specific conditions on Earth.
\end{abstract}


\section{Introduction}
Understanding how meaning is generated and represented has long been a central question in cognitive science and philosophy~\cite{burgess2014}. This concern, often referred to as the \textit{symbol grounding problem}~\cite{harnad1990} or the problem of \textit{metasemantics}, asks: How do abstract symbols, such as words, images, or neural activations, acquire meaning? Traditional approaches suggest that meaning is somehow grounded in sensorimotor experience, but the exact nature of this grounding remains controversial. In recent years, the emergence of high-dimensional embedding techniques across modalities for AI systems, such as neural signals, text, video, and audio, has opened a new avenue for investigating this problem. These methods allow us to represent diverse types of data as abstract numerical vectors, which in turn enable quantitative comparisons across modalities.

We investigate the following questions: Is there an underlying invariant semantic structure that is common to independently trained representation spaces (e.g., neural, textual, visual, and auditory modalities)? If such a structure exists, is it an inherent property of meaning itself? Or might it instead reflect the shared biological, environmental, and cultural conditions under which all our available data and methods are inevitably generated by interactions on Earth?

To set the stage, it is important to note that each modality's embedding space is learned independently. For example, neural embeddings are derived directly from fMRI data capturing brain activity, while text embeddings are produced using models such as BERT that are trained on vast corpora of language. These spaces are constructed with different objectives, architectures, and training data, and, by design, they start out with no inherent relationship to one another. They are abstract, high-dimensional representations that encode the information deemed important by their respective training processes. The issue is whether, despite these differences, a simple mapping or a comparative analysis reveals that these spaces converge on a shared semantic structure. Somehow, these distinct methods and datasets seem to end up generating the same way of representing the world. Is that really happening? And, if so, why?

What emerges from the studies surveyed is an intriguing possibility: there may exist a modality-independent essence of meaning itself, a relational structure that manifests consistently regardless of the particular form or modality through which it is expressed. Rather than simply reflecting arbitrary conventions, algorithmic presuppositions, superficial sensory properties, or brain organization, our representations might tap into a deeper semantic organization. The meaning we perceive might not be solely dictated by the specific features or sensory channels through which information is conveyed but by some underlying structures that remain remarkably stable across diverse domains of human experience.

In Section Two, we review and synthesize evidence from two broad families of methods. First, we discuss internal structure comparison methods that examine the geometry or relational patterns within each modality's space without learning an explicit mapping. Second, we examine transformation-based approaches that extract embeddings independently and then learn a transformation, often a simple linear mapping, to bring the spaces into correspondence. A successful alignment with a minimal transformation suggests that the spaces are nearly isomorphic, indicating that information is encoded in a similar way across modalities. Section Three contains an analysis of these results from the perspective of contemporary literature on the symbol grounding problem~\cite{harnad1990}, Bender and Koller's popular example of the statistician octopus~\cite{benderkoller2020}, and the Platonic Representation Hypothesis~\cite{huh2024}, which is the claim that semantic information encoded in diverse embeddings represents a fundamental level of reality, much like Plato's forms (on one interpretation). We reject the Platonic Representation Hypothesis and offer some reasons to doubt it, even if one accepts that there is an invariant semantic geometry across embedding modalities. Section Four outlines a number of challenges for thinking that an invariant semantic structure is truly universal. These challenges include the possibilities that the invariant structure comes from the embedding algorithms themselves, that it depends on how human brains process information, and that it is local to the human environments found on Earth. A brief conclusion includes some open questions and directions for future research.

\section{Methodological Approaches and Their Evidence}
An \textit{embedding} is a way of representing something, like a word, an image, or a sound, as a numerical vector (an ordered list of numbers). For example, instead of representing the word ``cat'' simply as letters, a language model might represent it as a list of numbers, such as $[0.1, -0.4, 0.5, \dots]$. A vector with \textit{n} entries can be thought of as a direction in \textit{n}-dimensional space; that is, one can visualize it as an arrow from the origin to the point whose coordinates are given by the list of numbers. These numerical vectors are designed so that the relationships between them correspond to meaningful relationships in the world: words with similar meanings, like ``cat'' and ``kitten,'' will be represented by vectors that are close together in an embedding space (i.e., they point in similar directions), while words with unrelated meanings, like ``cat'' and ``truck,'' will have vectors that are farther apart (i.e., their directions are very different). Because embedding spaces represent conceptual similarities as spatial relationships, they allow researchers to quantify semantic connections and systematically compare meanings across large sets of data. Thus, embedding spaces give structure to abstract concepts, turning complex information---such as the meanings of linguistic expressions, the content of images, or patterns of neural activity---into a mathematically organized representation. The overall structure of an embedding space is often called its \textit{geometry}, because it is characterized by geometric relations among vectors, such as distances and angles.

Because we are investigating the nature of vector embeddings across many modalities, like different languages, images, sounds, etc., we use the terms ``meaning'' and ``semantic'' in a more general way than is traditional in philosophical discussions. These terms are often used only for linguistic expressions or perhaps mental states, but it is less common to say that images have meanings or contain semantic information~\cite{HymanBantinaki2021}. However, we intend to use these terms to indicate general representational content, which includes what a word or image is about. Moreover, if the Representational Alignment Hypothesis is correct, then there is additional reason to extend categories like meaning and semantic content to anything that participates in the same invariant embedding geometry.

Modern AI models rely heavily on embeddings precisely because representing diverse data numerically makes it possible to detect and reason about complex relationships. Rather than working directly with raw inputs like pixels or text strings, these models first translate inputs into embedding vectors and then perform their analysis within the embedding space. The embedding spaces we examine are produced independently from one another by using fundamentally different data sources and training procedures. Text embeddings, for example, are derived from statistical patterns across vast textual datasets, whereas neural embeddings originate from recordings of brain activity. Because each modality's embedding space is generated separately, without reference to other modalities, investigating whether these independently produced spaces still share common structural patterns or converge can shed light on whether a deeper, possibly invariant semantic structure is captured by each of them. In other words, the key question is whether different embedding spaces---created from text, vision, audio, and neural data in isolation---nonetheless exhibit \textit{alignment} or similar internal geometry, suggesting a shared organization of meaning that goes beyond modality-specific constraints. Now we turn to the two major families of evidence for the RAH.

\subsection{Internal Structure Comparison Methods}
\textit{Internal structure comparison methods} assess the natural geometry and relational patterns within each modality's embedding space without first learning an explicit mapping between them. Instead of forcing one space to conform to another, these methods analyze how items relate to each other within each space. For example, representational similarity analysis (RSA)~\cite{kriegeskorte2008} computes a full pairwise similarity matrix (often using cosine similarity) for a set of items and then compares these matrices across modalities using correlation measures (e.g., Spearman's rho). Similarly, mutual information (MI)~\cite{shannon1948} estimation quantifies how much information one modality's representations share with those of another, while topological approaches~\cite{carlsson2009} examine the ``shape'' of the space by identifying clusters, loops, or other invariants. In essence, these techniques ask: Does the internal structure (the way items group together and relate) remain consistent across different types of data? Together, these methods provide a holistic, unsupervised perspective on whether an invariant semantic structure exists.

\subsubsection{Pairwise Similarity Analysis}
Many studies employ methods like RSA~\cite{kriegeskorte2008} to directly compare the relational patterns within each modality's embedding space~\cite{Cichy2014,aho2023,he2022}. By constructing similarity matrices for a fixed set of items, researchers can evaluate whether the patterns of distances and similarities among items are consistent across different spaces. High correlations between these matrices suggest that the modalities share a similar internal organization, lending support to the hypothesis of an underlying invariant semantic structure.

\begin{figure}[h]
    \centering
    \includegraphics[width=0.95\textwidth]{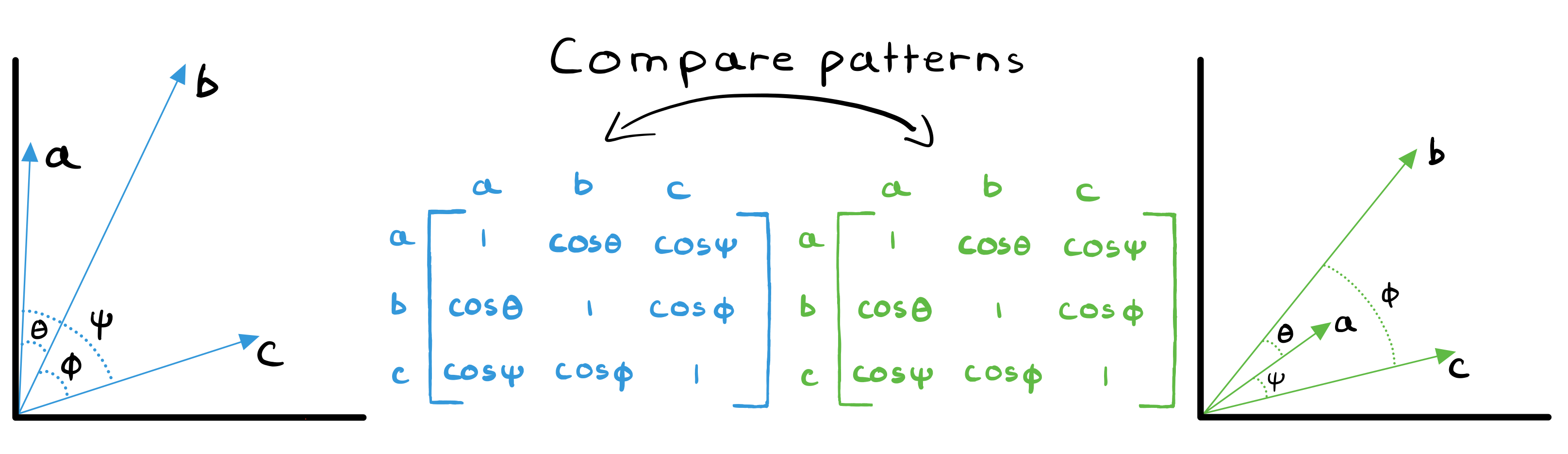}
    \caption{Representational Similarity Analysis (RSA) of two 2D vector spaces (green, blue), each containing three vectors. Similarity matrices are computed within each space and compared.}
\end{figure}

\subsubsection{Global Geometric and Topological Features}
In addition to pairwise comparisons, some approaches focus on the overall ``shape''~\cite{Singh2020,Kumar2015,Brown2025} of each embedding space. Techniques such as topological data analysis explore features like clusters, loops, and other invariants that characterize the global geometry of the data. When such topological features are consistent across modalities, it implies that even without an explicit mapping, the spaces exhibit comparable global structures. This reinforces the idea that the organization of meaning may be inherently shared across different forms of representation.

\subsubsection{Literature}
Aho et al.\ model early word learning as matching relational structure across modalities, even when direct word--object co-occurrences are unavailable or sparse~\cite{aho2023}. They employed GloVe-based embeddings (Global Vectors for Word Representation)~\cite{aho2023,pennington2014} for both image and word data, and then gave the agent a small set of concepts (word--image links). In each forced-choice trial (two novel words and two novel images), the agent evaluates the two possible matchings and chooses the one that makes the word-side similarity relations to the known concepts line up best with the image-side similarity relations to those same concepts. Accuracy was already $\sim$80\% with about 21 known concepts and rises to $\sim$90\% with 83 concepts, and learning these concepts in a childlike age-of-acquisition order beats random orders, suggesting that early-learned concepts provide a better relational scaffold for later cross-modal alignment~\cite{aho2023}.

Applying this principle directly to mature AI models, Schnaus et al.\ tackle the ``blind matching'' of vision and language models using a Quadratic Assignment Problem (QAP) to align their internal geometries without any paired data~\cite{Schnaus2025}. In a small-scale proof-of-concept on a 10-class dataset, where the optimal solution could be found by checking every possible combination, they achieved up to 100\% matching accuracy. To test a more scalable application, they used their proposed solver to create a fully unsupervised classifier, which reached 51.1\% accuracy on the CIFAR-10 benchmark, a result far exceeding the 10\% random baseline~\cite{Schnaus2025}. This work demonstrates that there is enough structural alignment between modalities to be exploited for practical, zero-shot supervision tasks.

A parallel thread was seen in neuroscience, where Cichy et al.\ analyzed magnetoencephalography (MEG) signals and fMRI responses in humans, alongside single-unit recordings from the macaque inferior temporal (IT) cortex, all for the same set of 92 object images, and constructed relational matrices for each: For the MEG data, they generated a decoding accuracy matrix using a machine learning classifier, while for the fMRI and monkey IT data, they created dissimilarity matrices calculated as ($1 - \text{correlation}$)~\cite{Cichy2014}. By comparing these matrices via representational similarity analysis (RSA)~\cite{kriegeskorte2008}, they showed that human MEG data initially aligned with V1-like representations ($\sim 101\,\mathrm{ms}$ post-stimulus) and later aligned with IT ($\sim 132\,\mathrm{ms}$)~\cite{Cichy2014}. Intriguingly, human MEG also correlated with macaque IT around 66--141~ms, suggesting a shared higher-level object representation across species~\cite{Cichy2014}. This fine-grained temporal matching underscores how similarity-based correlational analyses can reveal when and how neural data converge onto comparable semantic structures, without needing a strict one-to-one alignment of individual points across data types.

In more specialized domains, olfactory perception offers a striking case of mapping physicochemical structure to human descriptive language. Kumar et al.\ compiled thousands of odorants alongside their perceptual descriptors (``citrus,'' ``woody,'' etc.) and found that chemically similar molecules exhibit matching descriptor patterns~\cite{Kumar2015}. By constructing co-occurrence graphs for descriptors and sorting molecules by their chemical features, the authors observe a significant overlap between these two spaces, suggesting an implicit homomorphism relating physicochemical structure and perceptual labeling. Notably, a random forest classifier trained solely on chemical attributes can predict odor descriptor categories with Receiver Operating Characteristic scores, which quantify how accurately a model can discriminate between different categories (0.5 indicates random guessing, while 1.0 denotes perfect classification), of around 0.7--0.8, reinforcing the notion that internal relationships in each domain (chemical vs.\ perceptual) parallel each other.

Within human language processing, the similarity of internal embeddings for language and brain data was demonstrated by He et al.~\cite{he2022}. Here, event-related potentials (ERPs), a type of brain activity measured by EEG that directly reflects neural responses to specific events, were recorded while participants listened to stories; these data were converted into word-by-word similarity matrices and then correlated with similarity matrices from FastText and ELMo embeddings. FastText is an embedding technique that captures word meaning based purely based on statistical co-occurrences of words in large text corpora, whereas ELMo produces embeddings that dynamically adapt based on surrounding context in sentences. Both models yielded positive correlations, but ELMo, being context-sensitive, showed a higher average correlation ($\rho \sim 0.086$ vs.\ $\rho \sim 0.057$ for FastText) with the ERP patterns, implying that contextual embeddings align more closely with how the brain encodes semantic relationships. Pushing this model-to-brain comparison further, Ryskina et al.\ investigated whether language models align with specific brain regions that represent concepts across different sensory modalities~\cite{Ryskina2025}. To achieve this, they introduced a ``semantic consistency'' metric to identify brain areas that respond similarly to the same concepts whether it was presented as a sentence, a word cloud, or a picture. Using fMRI data, they located three such cross-modal regions and compared their representational geometry to that of 15 different language and vision-language models using representational similarity analysis (RSA). Their findings showed a significant alignment between the models and these semantically consistent brain areas, even in regions with low language sensitivity. These studies provide a real-world demonstration of how purely distance-based comparisons can reveal convergences in how machines and humans organize word meaning.

This convergence between model and brain representations is not only evident in passive neural recordings but also surfaces when comparing model geometry to active human cognitive judgments. For instance, Du et al.\ analyzed 4.7 million ``odd-one-out'' similarity judgments for 1,854 objects~\cite{Du2025}. These judgments were collected from human participants and generated by a large language model (ChatGPT-3.5) and a multimodal large language model (Gemini Pro Vision 1.0). Using a sparse positive similarity embedding (SPoSE) method, the authors derived a stable, low-dimensional (66-dimensional) representational space from the judgments for each group (humans, LLM, and MLLM). The dimensions of these model-derived spaces proved highly interpretable, reflecting semantic categories like ``animal-related'' and perceptual features like ``fine-grained pattern.'' The model embeddings showed strong alignment with both the human-derived embeddings and with neural activity patterns in category-selective brain regions, including the fusiform face area (FFA) and parahippocampal place area (PPA).

Across these varied contexts---including child concept learning, neural recordings, odor perception, and multimodal embeddings---a common theme emerges: evaluating how items internally cluster or correlate can reveal deep similarities (or differences) in representation. Whether through RSA, manifold alignment, topological signatures, or neighborhood graphs, these studies collectively uncover a shared geometry even in the absence of explicit point-to-point matching.

\subsection{Transformation-Based Methods}
Further evidence for the RAH comes from considering transformations between embedding spaces. A \textit{transformation}, in this context, is a mathematical operation that converts one set of embeddings into another, making the two sets directly comparable. The simplest and most common transformations used are linear mappings, which involve operations like rotating, scaling, or shifting the embedding vectors without significantly altering their internal relationships. If such straightforward linear operations are enough to align independently derived embeddings from different modalities, then the fundamental semantic relationships encoded by these embeddings are essentially the same, even though they originated separately. The fact that a simple mapping can align them suggests that the underlying structures are nearly isomorphic. That is, they share the same organization or geometry.

It is important to remember that the need for a transformation in no way diminishes the claim that the underlying relations are the same. For instance, if we were to take a single data set and train embeddings with the exact same algorithm twice, the two resulting embeddings would not actually be identical because each embedding space begins with a random assignment to every vector. That is, the vector for ``cat,'' for example, in the first embedding and the vector for ``cat'' in the second embedding would point in different directions. Nevertheless, all the vectors in the first and second embeddings would be related by a simple transformation. Essentially, the information about an item in a data set is encoded \textit{not} in the absolute direction of its vector, but rather in its relationship to all the other vectors, which is what remains invariant under transformations. The semantic information is contained in the difference between the direction of the vector in question and the directions of all the other vectors in the embedding~\cite{Gunther2019}. The information is relational, not absolute. Only in an embedding of many other vectors does a vector have any semantic information at all.

\subsubsection{Unsupervised or Minimally Supervised Alignment}
Many studies have found alignment with very limited external guidance~\cite{Roads2020}. For instance, some methods use a small bilingual dictionary, a modest list of known corresponding word pairs between languages, to provide minimal explicit examples that suggest an overall mapping between one representation space and another. Other studies use methods that operate entirely without supervision~\cite{Lample2018,Artetxe2018,Chung2018,Luo2024}, meaning they do not rely on any labeled examples or explicitly marked correspondences between modalities. Instead, these unsupervised methods must identify underlying similarities purely from the inherent patterns within the data itself. This minimal reliance on external input suggests that the observed correspondences between different modalities are not simply artifacts created by external intervention, but rather reflect genuine structure already present within the data.

\subsubsection{Linear vs. Nonlinear Mappings}
The literature distinguishes between linear transformations and more complex, nonlinear mappings. Linear methods, such as Procrustes alignment~\cite{gower1975}, which optimally match two embedding spaces by minimizing distances between corresponding vectors, are particularly compelling because they are simple and interpretable. If a basic linear transformation is sufficient to demonstrate alignment for two embedding spaces, it suggests that the differences between them are limited to basic geometric operations, such as rotation and scaling, rather than complex distortions. Nonlinear methods, while potentially uncovering stronger alignment, can obscure whether the shared structure is naturally present or artificially imposed.

\subsubsection{Literature}
A prominent focus in transformation-based methods has been cross-linguistic word embedding, where researchers attempt to map word representations from one language onto another. Early research, such as by Lample et al., demonstrated an unsupervised adversarial approach, where embeddings from one language are adjusted so that an automated discriminator (a model trained to distinguish embeddings from two languages) can no longer reliably distinguish source from target language embeddings~\cite{Lample2018}. Once the initial transformation is established, they apply a secondary refinement step with Procrustes analysis, which involves further adjusting embeddings with the help of an automatically generated ``synthetic dictionary.''

\begin{figure}[h]
    \centering
    \includegraphics[width=0.95\textwidth]{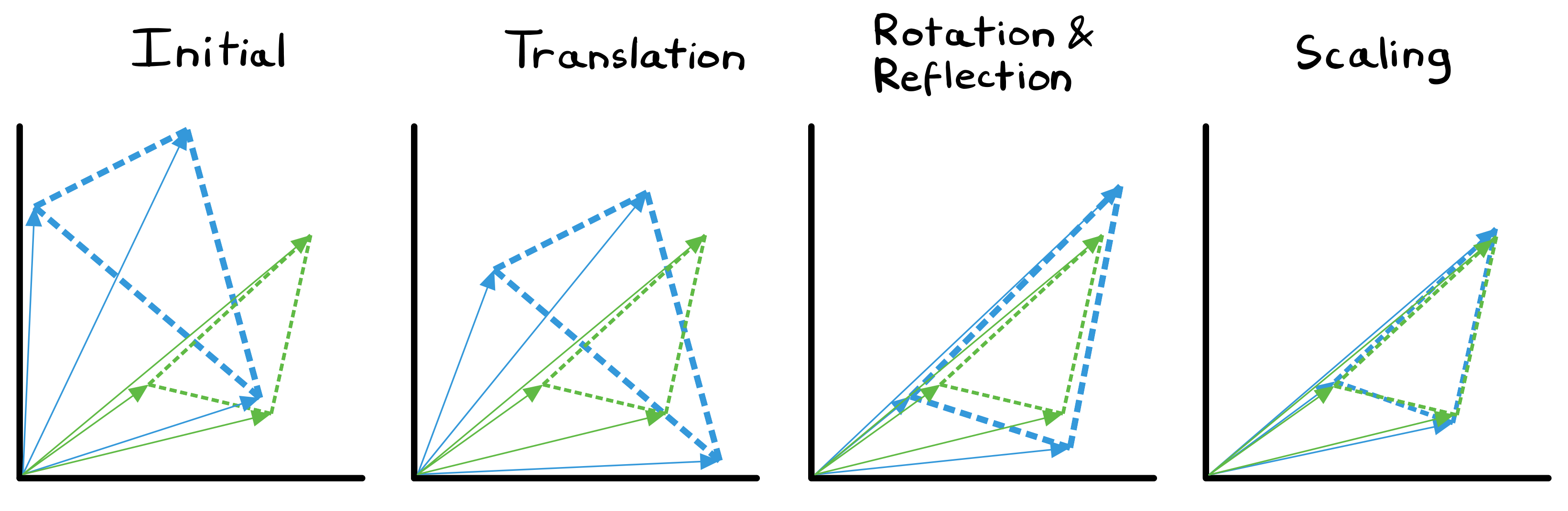}
    \caption{Procrustes analysis of two 2D embedding spaces (green, blue), each with three vectors. Solid lines represent vectors; dotted lines indicate ``shape'' of embedding space.}
\end{figure}

A critical innovation from this same work addressed the phenomenon known as the ``hubness'' problem~\cite{Lample2018}. In high-dimensional embedding spaces, the concept of distance becomes counterintuitive: rather than points spreading out evenly, some words end up being moderately close to many others simply by chance~\cite{Radovanovic2010}. As a result, these words become ``hubs'' that appear frequently as nearest neighbors, even though they are not genuinely similar or meaningfully related~\cite{Lample2018}. To mitigate this issue, the authors introduced Cross-Domain Similarity Local Scaling (CSLS), a method that rescales similarities between embeddings to balance out overly dense neighbors. Consequently, automatically generated word-for-word translations improve dramatically, even matching or surpassing supervised methods on common language pairs such as English--Italian or English--Spanish~\cite{Lample2018}.

Another line of work by Artetxe et al.\ first aligns embeddings by comparing similarity patterns across languages directly, using ordered distributions of similarities rather than relying on a discriminator~\cite{Artetxe2018}. This method starts with an initial seed mapping that is then iteratively fine-tuned using a straightforward, mathematically constrained transformation (an ``orthogonal transform,'' which preserves geometric relationships). This method effectively discovers an alignment even between highly dissimilar languages (for instance, English--Finnish), consistently avoiding failures that typically occur with more fragile adversarial setups. In practice, this stable method achieves results comparable to, and occasionally even exceeding, those of lightly supervised systems in bilingual lexicon extraction tasks~\cite{Artetxe2018}..

Pushing this tradition further, Jha et al.\ introduce \textit{vec2vec}, a method for performing direct, unsupervised translation of text embeddings from one model's space to another, without access to any paired data, the original text, or the source encoder model~\cite{jha2025}. Their approach, inspired by cycle-consistent adversarial networks~\cite{Zhu2017}, learns to map embeddings into a shared latent space and decode them into the target space~\cite{jha2025}. This technique proved highly effective, achieving cosine similarities as high as $\sim$0.92 between the translated vectors and their ground-truth counterparts in the target space. he authors further show
that these translated embeddings retain enough semantic information to enable downstream attribute inference and partial text reconstruction~\cite{jha2025}.

Researchers have also extended unsupervised alignment methods to speech--text modalities~\cite{Chung2018}. Chung et al.\, for example, learn a linear mapping between independently trained speech-based and text-based embeddings using an adversarial step plus a refinement phase akin to Procrustes. Remarkably, this alignment enables zero-shot spoken word classification: given a previously unseen audio segment, the model can retrieve the best-matching text word. While unsupervised alignment performance lags behind a fully supervised baseline, it remains impressively robust, especially for languages with limited parallel data. This result indicates that a relatively simple transform can bridge speech and text domains when their internal structures are sufficiently similar~\cite{Chung2018}.

Moving toward neural data, Goldstein et al.\ address whether the human inferior frontal gyrus (IFG) encodes words in a high-dimensional space that can be linearly mapped onto GPT--2 embeddings~\cite{Goldstein2024}. Using a stringent zero-shot split over non-overlapping \emph{unique} words (training on $\sim$90\% and testing on a held-out 10\%), the authors confirm that a learned linear mapping can accurately predict each new word's brain embedding. Even more, context-sensitive GPT--2 embeddings outperform static GloVe, suggesting the brain's IFG region encodes lexical items in a similarly ``dynamic'' embedding space~\cite{Goldstein2024}. In related work, Goldstein et al.\ also capture the neural basis of natural language processing in everyday conversations. Here, researchers align acoustic, speech, and language embeddings from a single model (Whisper) to electrocorticography (ECoG) signals, which are direct electrical recordings from the surface of the brain, collected over $\sim$100\,h of real-life conversation~\cite{Goldstein2025}. Linear regressions from each embedding layer to neural data reveal that speech embeddings align more strongly with lower-level auditory areas, while language embeddings align with higher-level language regions, underscoring how hierarchical alignment can reveal different cortical correspondences~\cite{Goldstein2025}.

Alignment also arises in conceptual systems that span text, images, and audio. Roads and Love argue that concepts have modality-specific embeddings whose role within each modality can be characterized relationally: each concept has a unique “signature” given by its pattern of pairwise similarities to other concepts, and this signature is “recapitulated” across modalities~\cite{Roads2020}. Using GloVe-based embeddings derived independently from text and from image/audio co-occurrence statistics, they score candidate cross-modal mappings by the Spearman correlation between within-modality similarity matrices under the mapping, showing that this alignment signal strengthens as the number of concepts grows—making unsupervised cross-modal correspondence increasingly easier in larger concept systems~\cite{Roads2020}. Furthermore, Luo et al.\ examine whether vision-only and language-only networks (trained independently) can be aligned without explicit parallel data~\cite{Luo2024}. Using Spearman correlations between within-modality similarity matrices (plus alignment strength and a recovery-accuracy measure), they find that self-supervised vision models (like DINO~\cite{caron2021} or iBOT~\cite{zhou2022}) exhibit stronger alignment to GloVe word embeddings~\cite{pennington2014} than purely supervised models do, presumably because self-supervision focuses on learning domain-general structures~\cite{Luo2024}. The success of such alignment underscores the possibility of isomorphic geometry even without direct image--text pairing, though a model explicitly trained on paired data remains the performance ceiling.

Extending these principles to the domain of socioeconomic analysis, Murugaboopathy et al.\ demonstrated their application in poverty mapping~\cite{Murugaboopathy2025}. The researchers aligned representations from two distinct, independently developed modalities: Landsat satellite imagery of neighborhoods and textual descriptions of these neighborhoods generated by a large language model's internal memory. To analyze cross-modal representational convergence, they align the resulting embedding spaces using Canonical Correlation Analysis, obtaining a median cosine similarity of approximately 0.60 after alignment. In prediction, combining vision and text substantially improves performance (the coefficient of determination, $R^2 \approx 0.77$) relative to a vision-only baseline (e.g., $R^2 \approx 0.63$).~\cite{Murugaboopathy2025}

In sum, these transformation-based studies provide compelling evidence that embeddings from different domains---including text, speech, images, and even neural signals---often share enough structural similarity to be reconciled by a simple learned transform. Across tasks as diverse as bilingual lexicon induction, zero-shot spoken word classification, conceptual alignment, and neural decoding, the successes of orthogonal mappings, adversarial and refinement loops, or direct correlation-based strategies all point toward a common underlying semantic geometry.

\section{Analysis}
Our review of cross-modal embeddings points toward a surprising and potentially profound conclusion: meaning is not arbitrarily assigned to the linguistic expressions of distinct languages, images, audio recordings, and other modalities, but instead displays an invariant structure across languages and modalities. The fact that relatively straightforward linear transformations suffice to align embedding spaces derived independently from neural, textual, auditory, and visual data strongly suggests that these modalities share an underlying organizational structure. Moreover, even when we move beyond explicit alignments and examine purely internal relationships, like how items cluster together and how their relational patterns overlap, this shared geometry remains robust.

The vector embeddings for each modality are independently derived without explicit reference to the others, arising from fundamentally distinct sources: linguistic patterns in textual corpora, pixel-level regularities in images, acoustic properties in audio signals, or neuronal activity dynamics in the brain. It is natural to expect meaning to be locally confined within the constraints and representational limits of each individual modality. Yet the evidence strongly suggests that meaning transcends these local boundaries, consistently converging onto a more invariant organizational structure. This is the conclusion we have dubbed the Representational Alignment Hypothesis (RAH). We have reviewed substantial empirical support for RAH, and it would be difficult to explain away all this evidence as mere coincidence. Nevertheless, the RAH remains a hypothesis because its full extent and robustness have yet to be conclusively investigated.

In this section, we suggest an explanation for why the RAH is true. Our explanation draws from research on the symbol grounding problem and contemporary metasemantics. Moreover, our explanation avoids the implausible metaphysical assumptions of competing explanations like the Platonic Representation Hypothesis. We begin with our reasons for rejecting the Platonic interpretation of the alignment. Then, in section 3.2, we discuss the symbol grounding problem as a framework and show how the RAH can help solve it. Further, in section 3.3, we consider Bender and Koller's recent statistical octopus example and explain the role the RAH has in undermining their argument. In the final section, we offer our own explanation for why the RAH might be true.

\subsection{A Platonic Reality?}
We have formulated the RAH as a general claim about vector embeddings for representations regardless of the modality, and we have offered an explanation for why it might be true and explored its significance for theoretical debates about symbol grounding. We want to distinguish the RAH from a view that might seem similar, but in our view, is unmotivated: the Platonic Representation Hypothesis (PRH). In this section, we explain the PRH, how it differs from the RAH, and why we reject the PRH.

Plato's Allegory of the Cave~\cite{plato1992} illustrates a fundamental tension between perceived reality and an underlying universal truth. In this allegory, prisoners who have spent their lives chained in a cave perceive only shadows projected on the wall, mistaking these mere representations for the full extent of reality. When a prisoner escapes and discovers the outside world, he recognizes that the shadows are just reflections of deeper, truer forms: a fundamental level of reality that transcends the limited experiences within the cave. The allegory emphasizes the distinction between surface-level perception and a potential universal, objective truth beyond human sensory limitations.

Recent work by Huh, Cheung, Wang, and Isola~\cite{huh2024} adapts Plato's allegory to representations in neural networks, observing that as models grow larger and more diverse, their internal embeddings tend to align. The proposed ``Platonic Representation Hypothesis'' suggests that these aligned embeddings reflect an underlying external fundamental truth, akin to Plato's forms. They write: ``Neural networks, trained with different objectives on different data and modalities, are converging to a shared statistical model of reality in their representation spaces''~\cite[p.~1]{huh2024}. They explain the connection to Plato's allegory as follows: ``the training data for our algorithms are shadows on the cave wall, yet, we hypothesize, models are recovering ever better representations of the actual world outside the cave''~\cite[p.~2]{huh2024}. In other words, as neural networks scale and gain broader experiences, they naturally uncover the same deep statistical structure of reality, more fundamental than the data used to train them, regardless of architectural nuances. Consequently, alignment across multiple modalities and architectures signifies not superficial coincidence, but a glimpse of a unified, universal reality that lies beyond the mere ``shadows'' of raw data.

Huh et al.\ offer the following elaboration on what they mean by the statistical model of reality that is analogous to Plato's forms in the allegory:

\begin{center}
  \begin{minipage}{0.8\linewidth} 
    \raggedright
 The world consists of a sequence of $T$ discrete events, denoted $\mathbf{Z}=[z_{1},\dots,z_{T}]$, sampled from some unknown distribution $P(\mathbf{Z})$. Each event can be observed in various ways. An observation is a bijective, deterministic function $\operatorname{obs}:\mathbf{Z}\to \ast$ that maps events to an arbitrary measurement space, such as pixels, sounds, mass, force, torque, words, etc.\ In this idealized world, knowing $P(\mathbf{Z})$ would be useful for many kinds of predictions; this would constitute a world model over events that cause our observations.~\cite[p.~7]{huh2024}.
  \end{minipage}
\end{center}

Notice that they assume that the world fundamentally consists of a sequence of events, but it is unclear what exactly they think the world model is. In this quote, it could be $P(\mathbf{Z})$ itself that constitutes the world model, or it could be knowing $P(\mathbf{Z})$ that is the world model. Neither of these options is very helpful, since $P(\mathbf{Z})$ itself is the distribution that produces the world (i.e., the sequence of events), and so it is not a model of the world. It is not a representation or a class of representations, so it is not the kind of thing to which our representations could be converging. Knowledge of $P(\mathbf{Z})$ is presumably a representation, but they say nothing about this knowledge, and so we are not any closer to understanding their overall hypothesis. Presumably, it can be internal to the neural networks they describe, but then we have to either say that neural networks literally have knowledge or accept that this is just a metaphor. Neither helps us get a clearer understanding of the overall hypothesis.

The authors go on to say that a world model is an internal representation, realized as a learned embedding space, whose dot-products reproduce the point-wise mutual information (PMI) of the latent world events: for two discrete events $a$ and $b$, $\operatorname{PMI}(a,b)=\log\!\bigl(P(a,b)/[P(a)P(b)]\bigr)$
quantifies how much more (or less) often the pair co-occurs than chance. Listing every event down the rows and again across the columns gives a symmetric \textit{PMI matrix} (also called a PMI \textit{kernel}) whose \((i,j)\) entry holds that value. Because the paper assumes each sensor is a deterministic bijection from latent events $\mathbf{Z}$ to observations, the PMI matrix computed in any modality is identical: it is a statistical skeleton of reality. Training an encoder with an InfoNCE-style or negative-sampling objective pushes its vector inner products to approximate those PMI entries. Thus, with unlimited data and model capacity, separate encoders (images, text, audio, or even different random initializations) \textit{converge} to embedding spaces whose dot-product kernels differ only by an orthogonal rotation. This shared PMI kernel constitutes ``knowledge of $P(\mathbf{Z})$''; possessing it is what the authors mean by having a world model, and the asymptotic alignment of independently trained embeddings on that kernel is what they mean by convergence.

There are three major differences between our RAH and their PRH. First, their view is that representations across machine learning models are converging over time: that is, the quantity of structure shared is increasing. However, our view is that there is a shared structure. We do not take a stand on the dynamics of this shared structure for this paper, but we agree that Huh et al.\ provide considerable evidence that the quantity of shared structure is actually increasing, i.e., the degree of alignment is increasing. Second, they seem to also think that the accuracy of the shared structure is increasing as well: ``If models are converging towards a more accurate representation of reality, we expect that alignment should correspond to improved performance on downstream tasks.''~\cite[p.~1]{huh2024}. We do not take a stand on this issue at all. We are squarely focused on the claim that there is a shared structure, \textit{not} how it might be changing over time. Third, their view is that the shared structure represents the fundamental level of reality or how the world is fundamentally. This is evident from their invocation of Plato's cave allegory. We, on the other hand, are focused entirely on the representations and the fact that languages and other modalities display an invariant structure. It is their metaphysical claim that these representations are getting at the fundamental level of reality that we reject, and we spend the rest of this subsection arguing against it.

Let us distinguish among the following theses:

\begin{enumerate}
  \item \textit{Representational Alignment Hypothesis (RAH)}: vector embeddings from distinct languages and other modalities display a common invariant structure. 
  \item \textit{Convergent Alignment Hypothesis (CAH)}: the common invariant structure across embeddings is growing in various ways. 
  \item \textit{Alignment Accuracy Hypothesis (AAH)}: the common invariant structure across embeddings accurately represents the world. 
  \item \textit{Increasing Accuracy Hypothesis (IAH)}: the accuracy of the common invariant structure across embeddings is increasing. 
  \item \textit{Platonic Representation Hypothesis (PRH)}: the common invariant structure across embeddings represents the fundamental structure of reality. 
\end{enumerate}

The relationships among these hypotheses are complex. Surely all the other four depend on RAH, but the converse is false, so RAH is the weakest. CAH depends on RAH, but on none of the others. AAH depends on RAH, but on none of the others. IAH depends on AAH, but not CAH, since the accuracy of the aligned representations could increase even if alignment is not converging. PRH surely depends on AAH, but not on CAH or IAH. The fundamental questions are: is there representational alignment at all? RAH says yes. Are the aligned representations accurate? AAH says yes. Are the aligned representations about the fundamental level of reality? PRH says yes. In addition, there are two other questions: is the amount of representational invariance growing? CAH says yes. And is the accuracy of the representations increasing? IAH says yes.

In this paper, we have focused only on RAH, whereas Huh et al.\ defend all five (without really distinguishing them) and focus on PRH. In the rest of this section, we argue against the PRH, and we do not take a stand on the other three.

Huh et al.\ offer three explicit explanations for why the Platonic Representation Hypothesis might be true:

\begin{enumerate}
  \item \textit{The Multitask Scaling Hypothesis}: There are fewer representations that are competent for N tasks than there are for M < N tasks. As we train more general models that solve more tasks at once, we should expect fewer solutions.~\cite[p.~6]{huh2024}
  \item \textit{The Capacity Hypothesis}: Bigger models are more likely to converge to a shared representation than smaller models.~\cite[p.~6]{huh2024}
  \item \textit{The Simplicity Bias Hypothesis}: Deep networks are biased toward finding simple fits to the data, and the bigger the model, the stronger the bias. Therefore, as models get bigger, we should expect convergence to a smaller solution space.~\cite[p.~7]{huh2024}
\end{enumerate}

Notice that all of these explanations focus on why the shared structure among languages and other modalities might be increasing (their topic), rather than explaining why there is any shared structure at all (our topic). It seems to us that if the authors were pressed on this issue, they might point to their formal result that different representation spaces share an identical representational structure, so long as they make some highly idealized assumptions about the nature of observation (i.e., that it is bijective). However, we do not agree that all texts, images, and items of other modalities ought to be thought of as observations. Plenty of writing does not encode observations about the world at all. The same goes for images created (e.g., just survey some of the plethora of AI-created images that have flooded social media in the past two years).

There are plenty of reasons to reject the PRH even though the RAH seems to be true. First, there are many linguistic expressions that obviously do not refer to anything at all (e.g., ``unicorn,'' ``Sherlock Holmes,'' and ``round square''). Many images (e.g., of a dragon) and audio clips (e.g., of a dragon's roar) are likewise not about the world in any way. There is, nevertheless, a shared structure they participate in. For example, the word ``dragon'' in English and the word ``dragão'' in Portuguese will each be assigned vectors in their respective embeddings that play an almost identical role in their respective embedding geometries (i.e., the vector for ``dragon'' will be similar to the vector for ``serpent'' and the vector for ``monster'' in the English embedding, while the vector for ``dragão'' will be similar to the vector for ``serpente'' and the vector for ``monstro'' in the Portuguese embedding). Images of dragons will be assigned vectors that are similar to vectors assigned to images of serpents and to images of monsters. None of this shared structure is explained by a Platonic metaphysical assumption, and it is not even possible to model it in the framework offered by Huh et al.

Second, many of our best explanations of the meanings of certain linguistic expressions are not compatible with the claim that they are about anything fundamental in reality. We urge the reader to try the experiment of placing one hand in warm water to heat up and the other hand in cold water to cool down before placing both hands simultaneously in room-temperature water and noticing that the room temperature water feels hot to one hand and cold to the other at the same time. These and other experiments suggest that many of the ways we describe reality are better understood as not picking out something \textit{objective} in the world but rather something \textit{subjective} in how we represent the world. Philosophers often refer to these as secondary qualities, following John Locke, who originally suggested the water temperature experiment~\cite{Locke1979}. Another example is that a claim like ``tomorrow will be hotter than today'' (assuming an interval scale like Celsius) is objective, whereas a claim like ``tomorrow will be twice as hot as today'' is not objective since it can be true on one scale (e.g., 5°C to 10°C) but false on a different scale (e.g., 41°F to 50°F). It might seem like the latter is describing reality, but this is an illusion. Similar phenomena can be found throughout language. In addition, some common words like ``the,'' ``and,'' and ``if'' do not seem to be about the world at all. Moreover, plenty of theorists argue that normative and moral terms like ``good'' and ``ought'' are also not about the world~\cite{Schroeder2010}. Finally, natural languages also display rampant vagueness, but few experts think that it is best explained in terms of metaphysics. Instead, many find epistemic or semantic explanations for vagueness that are more plausible~\cite{Sorenson2022}.

Third, even words that seem to be about some aspect of the world might not pick out anything fundamental. Of the theories of normativity or ethics that treat words like ``ought'' and ``good'' and ``right'' as if they represent something in the world, many assume that it is \textit{not} fundamental. Naturalist theories, in particular, tend to treat the entire normative realm as derivative~\cite{Lutz2024}. In Platonic terms, this would be something like the claim that there is no form of the good. Instead, goodness would be dependent on the real forms. Anti-realist theories abound in philosophy, and they surely would not take their subject matter to be fundamental~\cite{Joyce2022, Balaguer2025}. To say that the doctrine under question---everything referred to by whatever representations is part of the invariant structure is fundamental---is not popular or plausible would be an understatement. Almost everyone who takes a stand on this topic disagrees.

Finally, it might be that many of the categories and concepts we think are fundamental might not be. This can be illustrated by Goodman's new riddle of induction in which he defines the word ``grue'' in the following way~\cite{Goodman1983}. An object is \textit{grue} if and only if it is green and observed before the year 2050, or it is blue and not observed before the year 2050. We know that all observed emeralds are green, and we know that all observed emeralds are grue. However, we infer from these observations that all emeralds are green, but we do not infer that all emeralds are grue. Goodman's new riddle is to explain the difference in the two inference patterns. One natural answer is that ``grue'' has a complex disjunctive definition, whereas ``green'' does not, so it makes sense to think that ``grue'' would not figure in inductive inferences. However, Goodman's response was to define a new term ``bleen'' in the following way. An object is \textit{bleen} if and only if it is blue and observed before the year 2050, or it is green and not observed before the year 2050. Now we can define ``green'' in terms of ``grue'' and ``bleen'' so that ``green'' gets a disjunctive definition: an object is \textit{green} if and only if it is grue and observed before 2050, or it is bleen and not observed before the year 2050. Likewise, we can define ``blue'' in terms of ``grue'' and ``bleen'' in a similar way. The upshot for our purposes is that one might assume that many of the terms from natural language can be given straightforward definitions, but this depends on the other resources available. It might easily turn out that many of our words from natural language end up requiring ``grue-like'' definitions in terms of the elements that are fundamental to reality. It is a mistake to assume that finding an invariant structure across embeddings of different languages and modalities is evidence that our representations are getting at fundamental reality.

\subsection{Symbol Grounding Problem}
The symbol grounding problem, originally formulated by Harnad~\cite{harnad1990}, asks how abstract symbols (like words or images) come to have their meanings. It was originally conceived as a problem for what was at the time known as Strong AI, which is the claim that symbol-processing systems (e.g., a computer) with suitable programs could have mental capacities like the ability to understand language. Searle presented a classic objection to Strong AI in~\cite{Searle1980}, and Harnad's symbol grounding problem was directly influenced by it. This was before the point at which machine learning algorithms (which do not involve explicit symbols or programs) came to dominate the AI space in the 21st century. However, the symbol grounding problem has persisted as a worry about machine learning algorithms as well (see~\cite{MolloMilliere2023})---it can be put as: how can the outputs of large language models like the GPT line from OpenAI or the Claude line from Anthropic come to have meaning? If these models are outputting meaningful expressions of natural language, then how are the meanings of those outputs grounded in the world? Harnad suggested that the solution to the symbol grounding problem is that symbols derive meaning from their direct linkage to perception---meaning is grounded through immediate sensory interactions with the external world~\cite{harnad1990}. Since then, a huge number of solutions have appeared (see work by Taddeo and Floridi for an early survey~\cite{TaddeoFloridi2005}).

However, the Representational Alignment Hypothesis might point toward a different account of how words and other representational devices are grounded. Specifically, our review shows that independently produced embeddings---whether from language, vision, audio, or neural signals---consistently align, forming coherent semantic structures even in the absence of explicit pairings or direct sensory mappings between them. There is a straightforward explanation for why this happens. As Søgaard writes:

\begin{center}
  \begin{minipage}{0.8\linewidth} 
    \raggedright
  But why, you may ask, would language model vector spaces be isomorphic to representations of our physical, mental and social world? After all, language model vector spaces are induced merely from higher-order co-occurrence statistics. I think the answer is straight-forward: Words that are used together, tend to refer to things that, in our experience, occur together. When you tell someone about your recent hiking trip, you are likely to use words like \textit{mountain}, \textit{trail}, or \textit{camping}. Such words, as a consequence, end up close in the vector space of a language model, while being also intimately connected in our mental representations of the world. (Søgaard 443)
  \end{minipage}
\end{center}

As Søgaard points out, the patterns of objects, properties, and relations in the world that we experience show up in our written language as patterns of word co-occurrences. This is no coincidence, nor is it mere correlation. Word co-occurrence patterns are caused by the co-occurrence patterns in things. Of course, this causal chain is indirect and complex. The patterns in things are experienced by humans who then create the texts in question. Nevertheless, there is a clear causal relationship at play here. It is easy to see that the same explanation extends to the other kinds of embeddings relevant to the RAH. That is, pixel co-occurrence patterns in images reflect the patterns in the world as well. The same goes for sound co-occurrence patterns and brainwave co-occurrence patterns. All of these modalities display patterns that are reflections of the patterns in the world, which are experienced by the humans who create these texts, images, audio recordings, and other modalities. We see the same geometry across these different modalities because each one reflects the same structure in the world as it is experienced by us.

We have been using the term ``representation'' for the items that are assigned vectors in embeddings, regardless of the modality, so words, images, sounds, and brain activations are all representations. Then we might say that it is the relationship between representation co-occurrence patterns and worldly co-occurrence patterns that explains the RAH. If that is right, it offers a compelling solution to the symbol grounding problem. The words produced by large language models, which use vector embeddings to represent semantic information about the relevant linguistic expressions, have their meanings in virtue of the fact that the co-occurrence patterns among words are related to the co-occurrence patterns among worldly things like objects, properties, and relations. Of course, there is not the space in this paper to offer a convincing solution in detail to the symbol grounding problem, but what we have said so far ought to be enough to at least take seriously the idea that RAH could solve it in a novel way: having the right co-occurrence patterns that are caused in the right way is all that is required to ground the outputs of large language models and other natural language processing techniques based on machine learning algorithms with embeddings.

\subsection{The Statistician Octopus}
The Octopus thought experiment~\cite{benderkoller2020} was introduced as a contemporary adaptation of the classic symbol grounding problem~\cite{harnad1990}, updated specifically to critique contemporary statistical language models rather than earlier symbolic or rule-based systems. Whereas the original scenario questioned whether mere rule-based symbol manipulation could ground meaning, this updated formulation instead targets whether sophisticated statistical models, trained solely on textual data, can ever genuinely grasp semantic meaning. By framing the problem around statistical learners, the experiment directly challenges the idea that meaning and grounding can emerge purely from textual patterns without explicit sensory or experiential references.

In this scenario, a highly intelligent octopus, serving as a metaphor for an advanced statistical language model, is imagined to be at the bottom of the ocean and it taps into an underwater cable linking two isolated humans communicating from separate islands. Over time, the octopus becomes highly proficient at imitating and generating realistic human language by recognizing and exploiting statistical patterns within the textual data alone. However, despite this proficiency in reproducing human-like language, the thought experiment suggests the octopus never truly understands the semantic content or meaning behind the words it manipulates. This alleged lack of understanding arises from the octopus's absence of direct sensory or experiential grounding in the real world. For example, Bender and Koller argue that if a person on one of the islands sent instructions for a coconut catapult through the cable to those on the other island, the octopus would be unable to understand them since it has no experience of coconuts or catapults. Thus, the central claim of the thought experiment is that statistical language models, when trained purely on text without explicit referential links to sensory or external experience, are inherently incapable of genuinely comprehending the meanings of the linguistic symbols they so effectively manipulate.

This claim has recently been challenged by demonstrating that genuine semantic grounding can indeed emerge from purely textual statistical regularities~\cite{sogaard2023}. As we have seen, embedding spaces generated independently from raw text data (via a large language model) and from visual data (via a computer vision model) can align closely through straightforward linear transformations, even without explicit grounding signals. Specifically, word embeddings produced by language models can be precisely mapped onto embeddings derived from visual models, accurately retrieving visual concepts based only on textual representations. Remarkably, this alignment is demonstrated using minimal initial reference points, and it generalizes effectively to previously unseen concept pairings. These results provide strong evidence that textual embedding spaces inherently contain relational structures sufficiently rich to approximate sensory grounding, thus challenging the view that explicit experiential grounding is strictly necessary to establish semantic meaning. The result fits well with the RAH and the other evidence for it we surveyed above.

By looking at evidence that independently derived embedding spaces from distinct and varied modalities (neural recordings, text, speech, and visual data) we have argued that these spaces can indeed be brought into close alignment through minimal linear transformations. The success of transformation-based approaches and internal structure comparisons confirms that embedding spaces learned without explicit cross-modal supervision still converge upon a common semantic geometry. In other words, not only does the evidence we have discussed point towards the conclusion that purely statistical regularities from textual data alone can encode implicit grounding, but they also generalize this insight beyond text and vision, encompassing neural signals and auditory data as well. The findings we have reviewed thus provide additional empirical support for the argument that semantic meaning can be embedded within independently learned representations, further undermining the claim that genuine grounding requires explicit sensory or experiential supervision.

Although a full discussion is beyond the scope of this paper, it should be clear that the RAH provides a procedure by which the octopus could come to have information about what the island dwellers are talking about: if it knows what the embedding vectors for one language or modality refer to and it knows the transformations to other embeddings, then it could come to know what the vectors in the other embeddings refer to. For example, if it has access to a camera, then it could calculate a vector embedding for the set of images it creates. With enough time, it could try out lots of potential transformations from this embedding to the islanders' language embedding, which is called unsupervised cross-modal mapping. Once it can identify some of the words in the islanders' language via their vectors and the transformation it finds, then it can use the transformation from the islanders' language embedding to the image embedding to create an image of a coconut, even though it has never seen a coconut. It is through this sort of process that the octopus, by exploiting the Representation Alignment Hypothesis, could come to know what the words of the islanders' language refer to, even if it has never experienced those things.

\subsection{A Cycle of Representation}
Instead of appealing to Platonic metaphysics to explain the Representational Alignment Hypothesis, we offer an alternative explanation in terms of the process that creates representations. Philosophical views on this are called metasemantic theories~\cite{burgess2014}. They are distinguished in the following way: \textit{semantic} theories attribute or specify meanings for linguistic expressions of some language fragment (almost never for an entire language), in a compositional way, whereas \textit{metasemantic} theories explain why those linguistic expressions have those meanings. Semantic theories say what the expressions mean; metasemantic theories explain why the expressions have their meanings. Although specifying meanings is a hallmark of analytic philosophy, and still is common, it has also been taken up by linguists. This is part of a larger process by which topics that had been part of philosophy are outsourced as sciences. Metasemantics, however, is practiced primarily by philosophers and cognitive scientists. The explanation of the RAH we favor turns out to be a complex metasemantic theory. We do not have the space to defend or even elaborate on the full theory. Nevertheless, a sketch will help orient the reader to the philosophical significance of the Representational Alignment Hypothesis.

The bumper-sticker is that invariant semantic structure comes from languages and other modalities displaying the same co-occurrence patterns (for words, pixels, noises, etc.) and these are caused by the co-occurrence patterns in the experiences of the humans who create and maintain the languages and other modalities. There is invariant structure because languages and other modalities reflect the co-occurrence patterns in the way humans experience the world.

The longer story is that there is, presumably, a world humans inhabit, which also contains all the texts, images, audio, brain scans, and whatever other representations might be in the scope of the RAH. We are not going to presume anything about the world except that we causally interact with it and we experience it as having objects, properties, relations, events, facts, and other categories of stuff as well (e.g., fields). In fact, we create all these representational devices and we create them with certain patterns. As one can see in Figure 3, there might be all sorts of important conditions on what aspects of the world and which aspects of how we experience it might show up in language, images, audio, brain scans, etc. Perhaps we have to triangulate with other humans in order to create certain representations. In addition, there are representations in ourselves as well, by which we mean something not very controversial: our brains and nervous systems and sense organs have parts and those parts function so as to represent aspects of the world as we experience it. Perhaps predictive coding is required for some~\cite{Sprevak2024}; some might need self-locating attitudes~\cite{Titelbaum2022}. What is clear is that we perceive the world, think about what we perceive, make predictions about the effects of certain actions, perform some actions intentionally, perceive the effects of those actions, think about what we perceive, and so on. This endless cycle, which is the essence of each of our lives, no doubt plays an important role in the kinds of representations we can create. It is clear that some of our actions result in the creation of representations like texts and images; other representations (in our brains) are created without our deliberate activity.

\begin{figure}[h]
    \centering
    \includegraphics[width=0.95\textwidth]{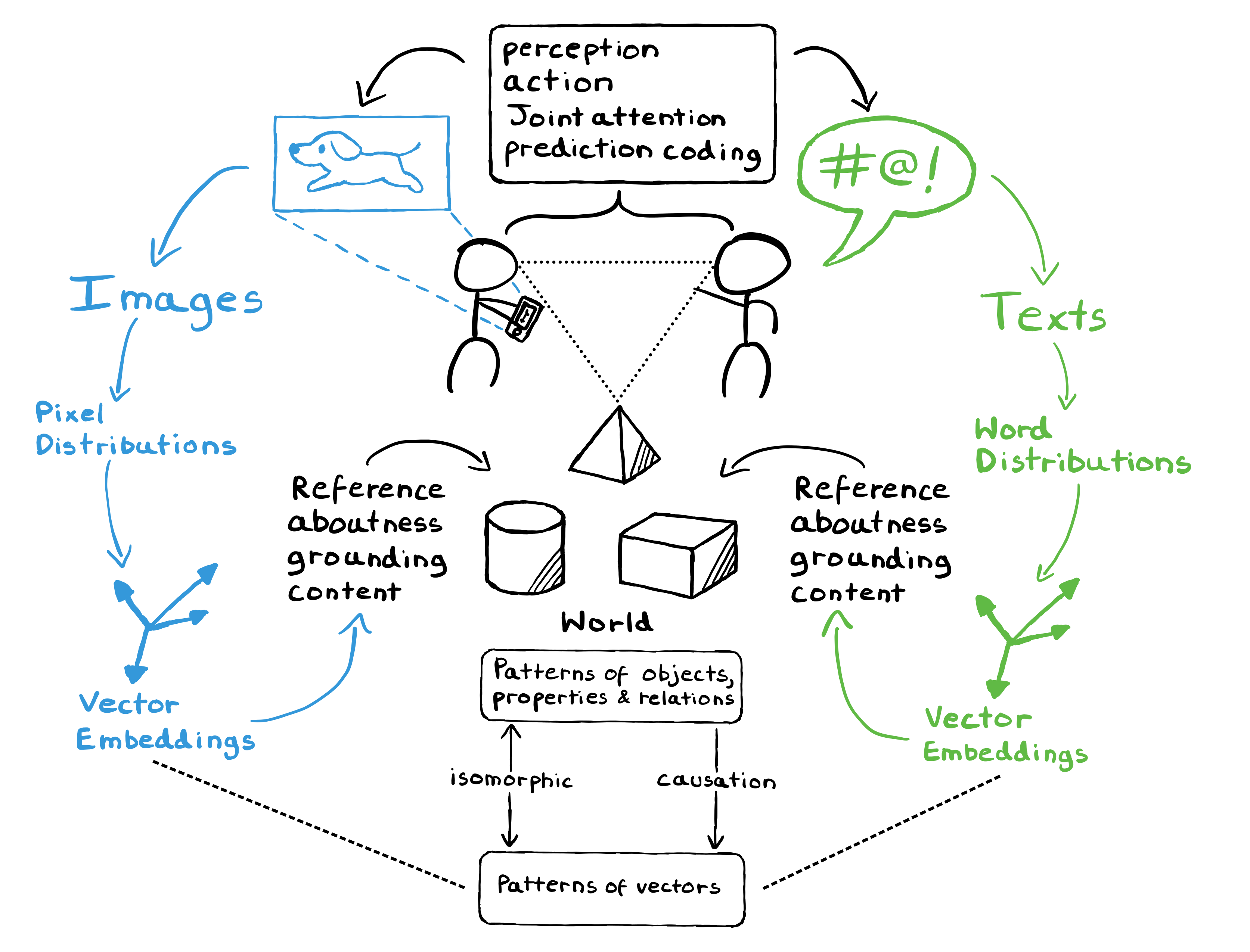}
    \caption{Causal-triangulation pathway from shared world events to convergent image--text embeddings.}
    
\end{figure}

The representations we create have parts like words and pixels. These parts show up in patterns. Of particular interest are co-occurrence patterns for these parts. The co-occurrence patterns are encoded in the vectors assigned to each part or representation. As in Figure 3, some of the representations, at least, are about the aspects of the world or items in the world. These items in the world are involved, perhaps indirectly, in the creation of the representations that are about them. That is why, on our account, those representations are about those items in the world. All it takes for representations to have their particular meanings is their involvement, in the right way, in this overall process. The shared invariant structure we have found across languages and other modalities is just encoded co-occurrence patterns in texts, images, etc. That invariant structure is caused by the fact that those representations were created by creatures who display certain co-occurrence patterns in their experience of the world. The way we interact with the world displays patterns that show up in our systems of representations. Overall, that's why our words mean what they mean, why our images depict what they depict, and so on. Notice that, according to this explanation of the RAH, not all elements of a language or other modality need to be about the world. They need only be part of a system of representations created in the process described.

For example, begin with two people attending to a common referent: a dog in the world. One person captures a photo with a smartphone, and the other provides a verbal description. Their shared focus on the dog sets the stage for triangulation: a speaker, a listener, and an object or event ``out there.'' The photo, the verbal utterance, and the physical dog each play a different role in anchoring meaning.

From that initial scene, the cycle splits into two mirrored pathways: one for the \textit{image-based} data, the other for \textit{textual} data. On the image side, we have the raw picture: color values, spatial arrangements, and all the other visible details that comprise the photograph. As the data pipeline continues, these pixels become distributions that capture statistical patterns (e.g., how frequently certain color intensities appear), ultimately mapping into high-dimensional \textit{vector embeddings} used in machine learning models. Meanwhile, on the textual side, the spoken words are transcribed or tokenized as text, then analyzed for co-occurrence frequencies, semantic similarities, and other linguistic regularities. In turn, they, too, collapse into vector embeddings. Each trajectory thus converts a real-world event, such as seeing a dog or commenting on it, into a numerical representation tuned for computational methods.

At the base of the diagram, the dog and its surroundings represent the ``patterns of objects, properties, and relations'' that exist independently of any representation. Aligned below them are ``patterns of vectors,'' the learned embeddings. Two lines connect these levels. The first is a bidirectional arrow suggesting a proposed \textit{structural isomorphism}: the hope that the relationships in the world can be mirrored in the geometry of the embedding space. This connection is complex. Somehow items in the world cause our experiences and our experiences give rise to our texts, images and so on. These serve as data for our embeddings. So the isomorphism between the embeddings and the items in the world is indirect and partial. The second line is an arrow of \textit{causation}, recognizing that the real-world entities, events, and interactions drive the raw data from which the model ultimately infers its vector patterns. These dual connections encapsulate both the philosophical aspiration that there is a meaningful one-to-one mapping between the ``way things are'' and ``the way embeddings represent them,'' and the pragmatic reality that data must first be generated by a physical cause (light bouncing off a dog, a person uttering words) before it ever becomes a numerical embedding.

Finally, the cycle closes as each embedding path circles back to the same world from which it was derived, maintaining the link of \textit{aboutness} or \textit{reference} at every stage. Rather than floating free as ungrounded symbols, the images and texts remain connected to the dog through the minds of the people who produce and consume these representations. At a higher level, this cyclical structure demonstrates why near-isomorphic embeddings might appear across modalities: all of them, however different in surface format, originate in the same external scene, are mediated by the same set of human interpretive acts, and converge into vector spaces that share an underlying geometry.

\section{Overarching Challenges}
Although these findings appear promising, several key interpretive and methodological issues remain. One central concern is the confounding variable problem: the possibility that the observed similarities between modalities might be explained by hidden or confounding factors, rather than indicating a genuinely invariant semantic structure. In the following subsections, we outline and discuss these critical challenges in greater detail.

\subsection{Lost in Embedding Translation}
A fundamental concern is that the embedding process itself may impose an artificial unity on data that would otherwise remain distinct. Many embedding methods, whether they are based on deep neural networks or more traditional dimensionality-reduction techniques, rely on common mathematical operations such as normalization, distance-based similarity measures, and standardized optimization objectives. By design, these methods tend to mold heterogeneous inputs into a somewhat uniform representation space~\cite{Bengio2013}. This can be compared to compressing different file types (e.g., text documents, images, and spreadsheets) into a ZIP archive: while the final compressed file is uniform in format, the original data might have very different structures, relationships, and content.

A similar pull toward uniformity can come from the training recipe itself. Training objectives like skip-gram with negative sampling effectively learn by factorizing a matrix of which words tend to appear together, so they can hard-wire a particular dot-product geometry into the space~\cite{Levy2014}. Modern contrastive losses (e.g., InfoNCE) further standardize the shape of embeddings by pulling matched examples together and spreading everything out on a unit hypersphere~\cite{Wang2020}, which can potentially make separately trained text, vision, or audio encoders easier to linearly map onto one another even without cross-modal supervision. Optimization might add another bias: Yu et al.\ find that models align more with lower \emph{local intrinsic dimensionality}, which is when their embeddings locally vary along only a few effective directions rather than filling the whole high-dimensional space~\cite{Yu2026}. Yu et al.\ further interpret the link between alignment and generalization through the lens of \emph{flat minima}: wide regions of the loss landscape where many parameter settings achieve nearly the same loss, so independent training runs can land at different points in the basin while remaining mutually alignable by a simple linear map~\cite{Yu2026,Hochreiter1997}. At the same time, these results also highlight why alignment can be fragile: if it arises from the geometry induced by a particular training setup, then changing that setup should change the geometry and reduce alignment. Indeed, small procedural changes (adding weight decay, transforming the targets, or introducing data heterogeneity) are predicted to attenuate or break it~\cite{Ziyin2025}. More broadly, this kind of fragility is consistent with work on implicit bias/implicit regularization, where the learned geometry depends on the optimization setup~\cite{Gunasekar2017,Soudry2018}.

Furthermore, the goal of unifying representations is undermined by the informational content of the data itself. In particular, there’s a critical distinction between tasks with redundant information and those requiring unique, complementary clues from each modality~\cite{Tjandrasuwita2025}. In such high-uniqueness settings, Tjandrasuwita et al. show that the alignment–performance relationship can weaken or even turn negative, and that increasing an explicit alignment objective can degrade performance when alignment is a poor predictor of task success~\cite{Tjandrasuwita2025}. Sarcasm detection is a canonical example where modality-specific cues, like tone and facial expressions, can change the interpretation of the text \cite{Castro2019}. In these cases, aggressively aligning representations can risk suppressing precisely the modality-specific signal needed for correct interpretation.

This concern that algorithm-driven uniformity may be misguided extends to the very techniques used to measure it. For instance, standard RSA can sometimes overstate alignment because it can be insensitive to a space's global topology~\cite{Brown2025}. By contrast, Representational Topology Analysis (RTA) probes higher-order structure (e.g., loops, clusters) and reveals cases where RSA reports similarity even as the underlying topologies diverge~\cite{Brown2025}. Beyond topology, widely used metrics like Centered Kernel Alignment (CKA) can yield misleading correspondence: because standard CKA is computed from centered kernel matrices and is invariant to orthogonal transformations of the feature space (including rotations and permutations of units)~\cite{Kornblith2019}, distorted scores can arise from shared anisotropy (common in contextual embedding spaces)~\cite{Ethayarajh2019}, outliers or dataset-level shifts~\cite{Davari2023}, or low-sample/high-dimensional bias in neural settings~\cite{Murphy2024} rather than genuine feature correspondence.

Ultimately, this highlights the critical need to ensure that any observed alignment is a genuine property of the data, not an artifact imposed by the analytical tools used to find it.

\subsection{Brains Behind the Bias}
Our day-to-day experience with data (digital photographs, audio recordings, text documents, neural scans, etc.) often feels like an objective glimpse of reality. We tend to assume that a photograph faithfully represents the world ``as it is,'' or that an audio clip captures sounds just as they exist. In truth, these forms of data are profoundly human-centered. Consider image data: standard RGB photographs capture only the band of visible light, having wavelengths from about 380 to 780 nanometres, that human eyes can detect, omitting the rest of the electromagnetic spectrum. Moreover, the camera's lens enforces a focus and depth of field that mimic our visual preferences, while compression algorithms (e.g., JPEG) discard ``unnecessary'' details in ways that humans typically do not notice. Similarly, our vantage point, usually at human eye level, reflects a field of view comfortable for our bodies. A species flying overhead or living deep underwater would capture images from radically different angles, use different color channels, and likely find our compression methods entirely inappropriate for its own perceptual priorities. Even audio data, which we often treat as ``merely recorded sound,'' is deliberately structured around human hearing ranges and loudness scales. Recording hardware and software are designed to accommodate the frequencies we can perceive, adopting sample rates (44.1 kHz, 48 kHz) optimized for our ears. Mono or stereo output is chosen to mirror the fact that humans have two ears, while temporal organization (clips, songs, speech segments) fits neatly into human attention spans and memory windows. In each case, the data sets we compile exist the way they do (bandwidth-limited to human needs, oriented in a human-friendly way) because we designed them around our own biological and cognitive constraints.

Ultimately, we use equipment built by humans for humans, employing color channels, sample rates, resolution standards, and labeling conventions that suit our perceptual faculties and interpretive capacities. Even neural data, which might appear to be an objective measurement of brain activity, is gathered using scanners (e.g., fMRI) or electrode grids (e.g., ECoG) designed to detect signals we consider relevant, filtered through assumptions about how and where neural activity becomes meaningful. Thus, although each dataset may seem to provide an unbiased ``raw'' record, in reality it reflects our vantage point: the wavelengths, frequencies, intensities, and scales we care about, and the forms of organization we impose to make sense of them.

From the perspective of the RAH, a deeper factor emerges: the brain's capacity to flexibly interpret any kind of input. Research on humans has shown that when a sensory modality is lost, the brain often reallocates that cortical territory in coherent, functional ways. For example, it has been demonstrated that in blind individuals, the visual cortex is recruited for tactile processing, such as reading Braille~\cite{sadato2005}. Similarly, it was shown that in deaf individuals, auditory regions respond to visual stimuli~\cite{finney2001}. These findings highlight the brain's cross-modal plasticity: its ability to repurpose regions based on available input rather than fixed modality-specific roles.

This suggests that our brains may have a general-purpose architecture that can efficiently reorganize around incoming data, regardless of how the data was originally ``intended'' to be processed~\cite{pascual2005}. Consequently, if we see that embeddings from text, images, and audio appear to converge or align, it may simply reflect the brain's unified strategy for encoding and making sense of any data it encounters. Rather than discovering a truly invariant semantic space, we could be observing how a single evolved intelligence, the human brain, tends to structure information of all kinds within its own constraints and biases~\cite{conway1968}.

\subsection{The View from Earth}
Another critical challenge to consider is that the semantic structure we identify might reflect constraints specific to Earth's unique biological and environmental context, rather than an invariant pattern applicable across the cosmos. All data we examine---whether textual, visual, auditory, or neural---is fundamentally generated within a narrow set of conditions tied directly to human existence on Earth. Our languages evolve from our interactions with the physical world around us, shaped by phenomena such as gravity, light, atmospheric composition, and terrestrial ecosystems. Similarly, sensory data like images and sounds inherently capture conditions and structures specific to our planet's environmental context (light intensities, gravitational orientations, acoustic frequencies, and more) shaping the semantic structure we derive from them.

To sharpen this point, imagine encountering an entirely alien form of intelligence from another planetary context. Would their sensory modalities, shaped by a vastly different environment, yield a semantic geometry remotely compatible with ours? If semantic representations are truly invariant in a cosmic sense, we might expect their embeddings to align structurally with ours despite radically different origins~\cite{eklund2024}. However, if the semantic structures we uncover are merely reflections of Earth's unique environmental and biological pressures, making them Earth-universal and universe-local rather than genuinely universe-universal, alignment across such fundamentally different origins could fail entirely.

\section{Conclusion}
Across diverse studies, two complementary research strands (transformation-based approaches and internal structure comparison methods) provide converging evidence that independently trained embeddings from text, vision, audio, and neural signals can exhibit a shared semantic organization.

A wide range of work has shown that simple transformations (often constrained to be linear or orthogonal) can align embeddings with minimal supervision. Such transformations have succeeded in cross-lingual word embedding alignment, speech--text mappings, neural decoding (e.g., aligning fMRI or ECoG signals with language-model embeddings), and even conceptual alignment between image and text spaces. In many cases, unsupervised or weakly supervised techniques (like adversarial training plus Procrustes refinement) yield performance close to or on par with fully supervised baselines, implying that the underlying geometry in these spaces is already ``near-isomorphic.''

Simultaneously, purely geometric or relational analyses (RSA, mutual information estimates, topological data analysis, and neighborhood-based similarity measures) have revealed strong cross-modal parallels in how items cluster or correlate. These studies span child concept learning (showing how visual and linguistic similarity alone can bootstrap new word mappings), cross-species neural alignment (human MEG correlating with monkey IT), single-cell multi-omics integration, and even odor perception.

Taken together, these two lines of evidence suggest that seemingly disparate modalities, trained under different objectives and data streams, frequently exhibit a coherent semantic geometry that can be revealed either by learning a simple mapping or by comparing their internal relational patterns.

A potential next step lies in constructing simulated environments where the underlying ``semantic'' factors are controlled and known~\cite{park2026}. By generating multiple modalities (e.g., visual frames, audio signals, or textual descriptions) from a shared virtual world, researchers can precisely measure whether independently trained models recover the true latent structure. Such an approach would enable systematic manipulation of variables, like adding noise in one modality or reducing data coverage in another, to observe how each factor affects the emergence (or breakdown) of an invariant representation. This setup would clarify whether near-isomorphism in embedding spaces stems from actual conceptual commonalities or merely coincidental correlations baked into real-world datasets.

Beyond simulation, cross-cultural and cross-species comparisons might offer another path toward understanding how deep these alignments really are. If embeddings trained on data from entirely different linguistic or cultural contexts still converge on similar relational patterns, it strengthens the claim that certain concepts or semantic dimensions transcend localized human experience. Conversely, finding persistent misalignments could reveal how much of this invariant structure depends on human-specific biology, history, or culture. Pushing this notion further, analyzing neural recordings or behavioral data from non-human primates, and comparing these with human-centric text or image models, would test whether meaningful alignment emerges when two systems inhabit distinctly different physical and cognitive realities.

Additionally, there is room for new interpretability methods that can reveal precisely which semantic attributes drive alignment across modalities. Even if two spaces appear correlated under standard metrics, it remains an open question whether they share the same conceptual axes---like animacy, shape, color, or more abstract relationships---or simply exhibit superficial similarity. Techniques like concept activation vectors, local neighborhood clustering, or topological data analysis could help identify whether consistent semantic groupings emerge in each modality and map onto each other in interpretable ways. By revealing the core dimensions along which embeddings align, such methods would help determine whether cross-modal convergence reflects a genuinely invariant geometry of meaning or a fortuitous byproduct of the ways these models are trained.

Finally, the machine learning landscape is shifting rapidly, and today's state-of-the-art multimodal systems differ markedly from those discussed here. We have not centered such models because most rely on paired cross-modal supervision that, by design, enforces alignment and would blur our focus on emergent alignment from independently trained modalities. Still, if future architectures or training schemes make it possible to isolate components that develop alignment without direct pairing, they could provide valuable evidence for emergence. If that proves difficult, then at the very least, SOTA systems can serve as positive-control calibration baselines: known-aligned references that verify our measurement pipeline, provide a ceiling for effect sizes, and help anchor results as the field continues to evolve.

\pagebreak
\bibliography{references}
\end{document}